\documentclass[aps, pra, superscriptaddress,twocolumn,showpacs]{revtex4-1}
\usepackage{graphicx}
\usepackage{xcolor}
\usepackage{amsmath}
\begin{document}
\title{Thermalization, condensate growth, and defect formation in an out-of-equilibrium Bose gas}

\author{D.~J.~Brown}\email{dylan.brown@auckland.ac.nz}\affiliation{Department of Physics, Dodd-Walls Centre for Photonic and Quantum Technologies, University of Auckland, Private Bag 92019, Auckland, New Zealand}
\author{A.~V.~H.~McPhail}\affiliation{Department of Physics, Dodd-Walls Centre for Photonic and Quantum Technologies, University of Auckland, Private Bag 92019, Auckland, New Zealand}
\author{D.~H.~White}\altaffiliation[Present address: ]{Department of Applied Physics, Waseda University, 3-4-1 Okubo, Shinjuku, Tokyo 169-8555, Japan}\affiliation{Department of Physics, Dodd-Walls Centre for Photonic and Quantum Technologies, University of Auckland, Private Bag 92019, Auckland, New Zealand}
\author{D.~Baillie}\affiliation{Department of Physics, Dodd-Walls Centre for Photonic and Quantum Technologies, University of Otago, Dunedin, New Zealand}
\author{S.~K.~Ruddell}\altaffiliation[Present address: ]{Department of Applied Physics, Waseda University, 3-4-1 Okubo, Shinjuku, Tokyo 169-8555, Japan}\affiliation{Department of Physics, Dodd-Walls Centre for Photonic and Quantum Technologies, University of Auckland, Private Bag 92019, Auckland, New Zealand}
\author{M.~D.~Hoogerland}\affiliation{Department of Physics, Dodd-Walls Centre for Photonic and Quantum Technologies, University of Auckland, Private Bag 92019, Auckland, New Zealand}

\begin{abstract}
\noindent We experimentally and numerically investigate thermalization processes of a trapped $^{87}$Rb Bose gas, initially prepared in a non-equilibrium state through partial Bragg diffraction of a Bose-Einstein condensate (BEC). The system evolves in a Gaussian potential, where we observe the destruction of the BEC due to collisions, and subsequent growth of a new condensed fraction in an oscillating reference frame. Furthermore, we occasionally observe the presence of defects, which we identify as gray solitons. We simulate the evolution of our system using the truncated Wigner method and compare the outcomes with our experimental results.
\end{abstract}
\maketitle

\section{Introduction}

\noindent Thermalization processes present within weakly interacting Bose gases play a crucial role in the experimental realization of a Bose-Einstein condensate (BEC) through forced evaporative cooling~\cite{Anderson1995}. 
Detailed modeling of dynamical processes within these confined Bose gas systems presents a challenging problem. While the dynamics of a pure BEC at zero temperature are well represented by a mean-field description~\cite{Leggett2001}, the addition of non-condensed atoms greatly increases the dynamical complexity of the system, and accurate simulation quickly becomes intractable~\cite{Lee2016}. To develop a deep understanding of the thermalization process, careful consideration of both theoretical models and experiments are required.

Early thermalization experiments involving ultracold Bose gases were concerned with the feasibility of evaporatively cooling to Bose-Einstein condensation. To this end, the elastic scattering properties and cross-dimensional mixing rates of atoms within a magnetic trap were investigated~\cite{Monroe1993,Wu1996}. 
With the experimental realization of a BEC~\cite{Anderson1995}, the growth rate of a condensed fraction from thermal vapor was observed~\cite{Miesner1998, Kohl2002}, and found to differ from a pure relaxation process. 
The quantum mechanical nature of cold atom scattering has been revealed through energetic BEC collisions~\cite{Thomas2004, Kjaergaard2004, Mellish2007}. Thermalization processes depend largely on the microscopic properties of the system involved, such as the $s$-wave scattering length. Dimensionality also plays a critical role in defining these processes, and work has been done to investigate the properties of ultracold Bose gases in one-~\cite{Kinoshita2006, Hofferberth2007, Gring2012}, two-~\cite{Ermann2016}, and three-dimensions~\cite{Lobser2015, Lee2016}. 
Despite this, there is currently no universally accepted theoretical description detailing the full growth, relaxation, and thermalization properties of ultracold atomic gases~\cite{Lee2016}.

In this paper we experimentally and numerically investigate the thermalization processes of an ultracold Bose gas initially prepared in a highly non-equilibrium state, formed by stimulating atoms from a $^{87}$Rb Bose-Einstein condensate into a higher momentum state through Bragg diffraction. The system is allowed to evolve within a three-dimensional confining potential, in which collisions occur as the atoms oscillate within the potential. The BEC is quickly destroyed and the atoms are scattered into various other modes, from which we observe the growth and reformation of a new condensed fraction in an oscillating reference frame. We model the initial growth of atomic density in this frame accounting for Bosonic enhancement~\cite{Miesner1998}, and compare growth rates for different harmonic trapping frequencies.
We also compare the time evolution of our system to simulations using the truncated Wigner method, and discuss the use of this method for modeling the evolution of such systems. 
Finally, in a fraction of experiments, we observe the formation of defects within the condensed fraction that we identify as gray solitons. We attribute these defects to the inability for phase information to quickly propagate as separate condensed fractions grow from different nucleation centers within the system, a process known as the Kibble-Zurek mechanism~\cite{Zurek1985,Zurek1996,Zurek2009}. 

\section{The Experiment}\label{sec:experiment}
\noindent Our experiment involves a BEC of approximately 2$\times10^4$ $^{87}$Rb atoms optically pumped into the $\left|F=1;m_F=-1\right>$ state, and held in an optical dipole trap~\cite{Wenas2008}. The dipole trap is formed at the center of two intersecting, focused CO$_2$ laser beams, with a wavelength of 10.6~$\mu$m, and each with a $1/e^2$ radius of 35~$\mu$m. 
This potential can be considered near harmonic around the trap center, and can be characterized by a set of frequencies $\omega_j$ that define the potential in all three dimensions. We determine these frequencies experimentally through a parametric heating process~\cite{Friebel1998}. 
Following the loading of atoms from a magneto-optical trap into the optical dipole trap, we perform a 6.5-s evaporative cooling sequence to produce a BEC.

\begin{figure}[t]
	\centering
	\includegraphics{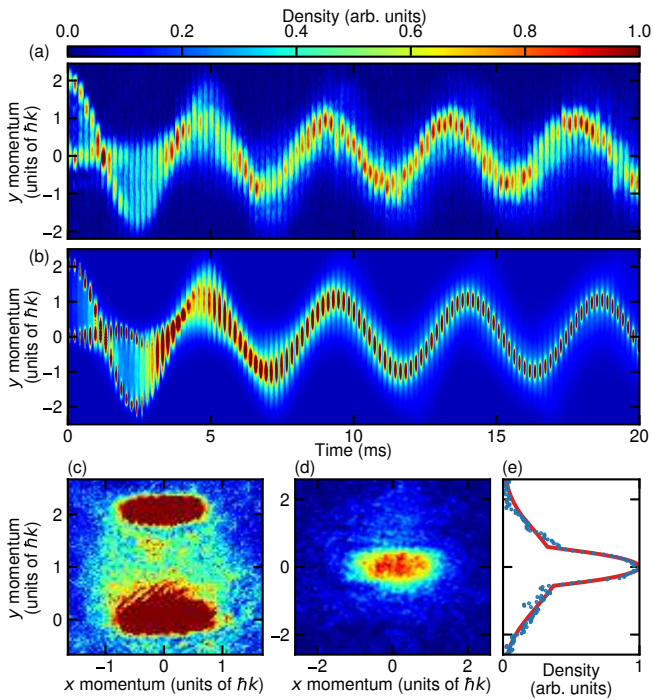}
	\caption{(a) A sequence of concatenated time-of-flight absorption images showing the momentum evolution of the system in the $y$-dimension as a function of in-trap hold time $t$. Each image is an average of three experimental realizations. (b) A numerical simulation of the same system evolution using the truncated Wigner method, as described in Sec.~\ref{sec:simulation}. (c) The average of 100 experimental time-of-flight images for $t=0$, with an $s$-wave scattering halo containing $\sim 25\%$ of the atoms formed. (d) A time-of-flight image following 16-ms of in potential evolution, where the system has thermalized to form a new condensed fraction. (e) Integrating along the $x$-dimension of (d) clearly shows the bimodal nature of the system. Here, we fit a Thomas-Fermi profile to the condensed fraction and a Bose-Enhanced Gaussian to the remaining atoms.}

	\label{fig:experimentEvolution}
\end{figure}

Once a BEC has been obtained, we 
place the system out of equilibrium as outlined below. Initially, the CO$_2$ laser power is adiabatically ramped up over 100~ms, resulting in an increased trap depth, which prevents atom loss when the sample is heated. This also allows us to have control over the trap frequencies. For the experiments presented here, we utilize trap frequencies in the range $\omega_y/(2\pi) = 150$--400~Hz, with $\omega_x = 1.2\omega_y$ and $\omega_z = 2\omega_y$, and with the $z$-dimension parallel to gravity. A 60~$\mu$s Bragg diffraction pulse~\cite{Stenger1999} is then applied in the $y$-dimension to coherently transfer 50\% of the BEC into an ensemble having $2\hbar k_\ell\hat y$ momentum, where $k_\ell = 2\pi/\lambda$ is the wavenumber of the Bragg diffraction laser. This laser has a wavelength $\lambda=783$~nm, detuned to the red of the D$_2$ resonance by $\Delta=$~1.6~THz to prevent heating by spontaneous emission within the atomic sample. 

The dipole trapping potential remains on throughout this process, and the atoms are allowed to oscillate within the potential, where they collide and rethermalize. Following a variable evolution time $t$ of up to 20~ms within the potential, the atoms are released from the potential, where they expand freely for 10~ms and are imaged using an absorption technique, yielding a time-of-flight image of the momentum distribution. This process is repeated for $t$ increasing in 200~$\mu$s steps, allowing us to construct a sequence of images representing the evolution of the system as it thermalizes, as shown in Fig.~\ref{fig:experimentEvolution}(a). For comparison, we also present a numerical simulation of the system evolution using the truncated Wigner method in Fig.~\ref{fig:experimentEvolution}(b), where the potential is Gaussian with $(\omega_x,\omega_y,\omega_z)/(2\pi) = (264,220,440)$~Hz and a $1/e^2$ radius of 35~$\mu$m. Further details of the simulation are presented in Sec.~\ref{sec:simulation}.

Care must be taken when interpreting time-of-flight images such as those presented in Fig.~\ref{fig:experimentEvolution}(a), as these images represent the momentum of the atoms, and hence lack information about the position of the atoms within the potential. As an example, the left-most image in Fig.~\ref{fig:experimentEvolution}(a) at $t=0$ shows that there exists two separate momentum ensembles, however, in-trap both of these components are initially spatially overlapped. This results in immediate collisions between atoms with $0$ and $2\hbar k_\ell \hat y$ momenta, which are then scattered into a range of momentum states. In Fig.~\ref{fig:experimentEvolution}(c), we show an average of 100 time-of-flight images of the system following the Bragg diffraction process, where we have observe the presence of a low density scattering halo~\cite{Gibble1995, Chikkatur2000} containing approximately 25\% of the atoms, consistent with an $s$-wave scattering description of the collision process. This provides the initial condition for the subsequent time evolution and rethermalization processes of the system within the potential.

\section{Thermalization}\label{sec:thermalization}

\noindent 
Following Bragg diffraction of the initial BEC, the majority of the atoms reside in two ensembles centered on two time-dependent momentum values, which we label as $|0\hbar k_\ell\rangle$ and $|2\hbar k_\ell\rangle$. Here, the momentum label refers to the  momentum amplitude of each ensemble, as the actual momentum oscillates due to the confining potential.  
Immediately following the creation of the two momentum ensembles, they begin to spatially separate within the potential. After evolving for $t\approx1.2$~ms, atoms in the $|2\hbar k_\ell\rangle$ ensemble reach the turning point of the potential, where they have a momentum close to zero. Here, the two ensembles have a spatial separation of approximately $10~\mu$m, whereas the spatial extent of each ensemble in the potential is about $1~\mu$m. Atoms in the $|2\hbar k_\ell\rangle$ ensemble then accelerate towards the center of the potential, where they collide with atoms in the $|0\hbar k_\ell\rangle$ ensemble starting from $t\approx2.0$~ms. As seen in Fig.~\ref{fig:experimentEvolution}(a), the remaining condensed fraction within each momentum ensemble is destroyed, and atoms are scattered into a range of momentum states.

From here, the system quickly coalesces into an ensemble oscillating at the average momentum of $|1\hbar k_\ell\rangle$, where we observe the growth of a new condensed fraction. Convergence of the system to a momentum amplitude of $|1\hbar k_\ell\rangle$ is to be expected, as this corresponds to the center-of-momentum of the system. However growth of a condensed fraction in this non-inertial reference frame is initially surprising, and demonstrates the ability of the system to develop and maintain phase coherence as the atoms are accelerated within the potential.

The $|1\hbar k_\ell\rangle$ state can be traced by considering the center-of-momentum of the entire system as a function of hold time. We observe a co-sinusoidal oscillation, from which we are able to extract the $y$-dimension trapping frequency as $\omega_y/(2\pi) = 220$~Hz. Although we expect that in a truly harmonic potential this center-of-mass oscillation would persist indefinitely~\cite{Dobson1994}, we observe a weak decay in the amplitude of the oscillation, with a time constant of 60~ms. We attribute this to anharmonicities within the trapping potential~\cite{Pantel2012}. Atoms with higher initial momentum travel further within the potential, and due to the Gaussian profile of the trap, experience a lower effective trapping frequency. These atoms therefore have a longer oscillation period within the trap, and the system will begin to experience an overall dephasing between different momentum components. This allows energy initially locked in the center-of-mass oscillation of the system to become available for thermalization.

Following approximately 3.5 oscillation periods within the potential, corresponding to $t\approx16$~ms, we observe that the system consists of a condensed fraction sitting on top of an approximately thermal fraction, oscillating with the center-of-momentum amplitude $|1\hbar k_\ell\rangle$, as shown in Fig.~\ref{fig:experimentEvolution}(d). We also show the integrated the atomic density along the $x$-dimension in Fig.~\ref{fig:experimentEvolution}(e) to emphasize the bimodal nature of the system following rethermalization.

\section{Experimental analysis}\label{sec:analysis}

\noindent To analyze the evolution of the system shown in Fig.~\ref{fig:experimentEvolution}(a), we first transform to the $|1\hbar k_\ell\rangle$ center-of-momentum reference frame, which oscillates co-sinusoidally with the trap frequency $\omega_y$. Growth of atomic density within this $|1\hbar k_\ell\rangle$ momentum amplitude arises through scattering of atoms from other momentum states. Prior to the Bragg diffraction process, the initial BEC has close to zero energy compared to the energy scale of our experiment. The total energy of the system, where half of the atoms are promoted to a momentum of $2\hbar k_\ell \hat y$, is $E_{\rm tot} = N\hbar^2k_\ell^2/m$, where $N$ is the total number of atoms in the system, and $m$ the mass of a single atom. As the system converges to the $|1\hbar k_\ell\rangle$ state, some of this energy becomes locked in a center-of-mass oscillation of the system, amounting to $E_{\rm COM} = N\hbar^2k_\ell^2/(2m)$, corresponding to half of the total energy. The remaining energy, $E_{\rm th} = E_{\rm tot} - E_{\rm COM}$, is available for thermalization processes, and contributes to heating. If $E_{\rm th}$ is less than the energy required to exceed the critical temperature for Bose-Einstein condensation $T_c$, growth of a condensed fraction oscillating in the center-of-momentum frame will be observed. While the one-dimensional scattering of two atoms, one from each of the $|0\hbar k_\ell\rangle$ and $|2\hbar k_\ell\rangle$ states into the $|1\hbar k_\ell\rangle$ state, conserves momentum, this process is not directly permitted due to energy conservation laws without either exciting the atoms to higher thermal states or transferring energy into other dimensions through cross-thermalization. Therefore, any condensed fraction that forms in the $|1\hbar k_\ell\rangle$ reference frame must grow from an ensemble of previously scattered atoms.

\begin{figure}[t] \centering

	\includegraphics{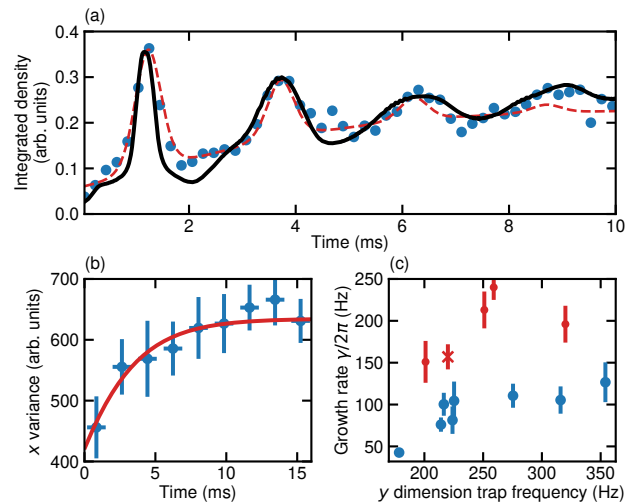}
	\caption{(a) Atomic density integrated within a $0.45\hbar k_\ell$ region about the $|1\hbar k_\ell\rangle$ center-of-momentum state as a function of in-potential evolution time $t$ for the experimental images shown in Fig.~\ref{fig:experimentEvolution}(a) (blue circles). We model this growth accounting for Bose-enhancement (red dashed line), from which we are able to extract a characteristic growth rate of $\gamma/(2\pi)=160\pm15$~Hz. A similar integration is performed for the numerical simulation previously shown in Fig.~\ref{fig:experimentEvolution}(b) (black line). (b) Growth of the momentum variance in the $x$-dimension indicating the presence of cross-dimensional thermalization, fit with a growth curve from which we find that $\gamma/(2\pi)=115\pm20$~Hz. (c) Growth rates of the $|1\hbar k_\ell\rangle$ state extracted for various systems which thermalize above the critical temperature $T_c$ (blue dots). The growth rates for systems that thermalize below $T_c$ are shown in red, with the data from (a) shown as a cross.}
	\label{fig:convFit}
\end{figure}

By integrating the atomic density within a $0.45\hbar k_\ell$ region around the center-of-momentum, we are able to observe the rate of growth of this atom number as a function of in-trap evolution time $t$, as shown in Fig.~\ref{fig:convFit}(a). We observe that atomic density within this region grows as atoms are scattered into the $|1\hbar k_\ell\rangle$ ensemble. The temporal density peaks correspond to atoms in both the $|0\hbar k_\ell\rangle$ and $|2\hbar k_\ell\rangle$ momentum ensembles overlapping with the $|1\hbar k_\ell\rangle$ ensemble in momentum space. Although the presence of these peaks partially obscures the growth of atomic density in the $|1\hbar k_\ell\rangle$ ensemble, their decay provides information about the depletion of atoms from both the $|0\hbar k_\ell\rangle$ and $|2\hbar k_\ell\rangle$ ensembles, which we expect is directly related to the growth rate of the $|1\hbar k_\ell\rangle$ ensemble. Additionally, these momentum-space peaks occur when atoms in the $|0\hbar k_\ell\rangle$ and $|2\hbar k_\ell\rangle$ ensembles have their maximal spatial separation, and are therefore not expected to impact the growth of the $|1\hbar k_\ell\rangle$ ensemble at this time. We therefore include these density peaks in our model of the growth rate of the $|1\hbar k_\ell\rangle$ ensemble. %as shown by the red dashed curve in Fig.~\ref{fig:convFit}(a).

For the data presented in Fig.~\ref{fig:convFit}(a), the energy available for thermalization, $E_{\rm th}$,  is not sufficient to exceed $T_c$, and hence we observe the formation of a condensed fraction of atoms in the $|1\hbar k_\ell\rangle$ center-of-momentum reference frame, along with the growth of a thermal state. We model this growth using a curve accounting for Bose enhancement within the system~\cite{Miesner1998}, allowing us to extract a characteristic growth rate $\gamma$ for the system. As the temporal peaks representing atoms in the $|0\hbar k_\ell\rangle$ and $|2\hbar k_\ell\rangle$ states passing through the integration region decay as the system evolves, we set this rate of decay to be equal to the growth rate of the atomic density in the $|1\hbar k_\ell\rangle$ state. This analytical fit to the data is shown in Fig.~\ref{fig:convFit}(a) as the dashed red curve, from which we extract a growth rate of $\gamma/(2\pi) = 160\pm15$~Hz. The black curve in Fig.~\ref{fig:convFit}(a) is from an integration about the centre-of-momentum reference frame for 50 runs of a truncated Wigner simulation, as discussed in Sec.~\ref{sec:simulation}. 

Due to the three-dimensional nature of our trapping potential, energy from the Bragg diffraction process initially applied in the $y$-dimension is distributed between all three dimensions as the system equilibrates. Figure~\ref{fig:convFit}(b) shows the increase of momentum variance in the $x$-dimension as a function of $t$, giving insight into the cross-dimensional thermalization rate. We fit a similar growth curve to the system, and find that the growth of the variance is $\gamma/(2\pi)=115\pm20$~Hz. 

Following approximately 16~ms of in-trap evolution, growth in the center-of-momentum reference frame has ceased in the $y$-dimension, and the system has approximately thermalized. Although the system will continue to evolve within the potential we are able to extract an instantaneous temperature of the ensemble, giving insight into how the system has thermalized. To achieve this we perform a bimodal fit of a Thomas-Fermi profile for the condensed fraction, and a Bose-enhanced Gaussian for the remaining atoms, as shown in Fig.~\ref{fig:experimentEvolution}(e), obtaining the atom number in each. Considering the condensed fraction of atoms we obtain $T=295\pm8$~nK, below the critical temperature $T_c=360\pm5$~nK for our system.
  
After 16~ms of in-trap evolution, we find that the momentum amplitude of the $|1\hbar k_\ell\rangle$ center-of-momentum state has decayed to $0.75\hbar k_\ell$, and hence $40\%$ of the initial $E_{\rm COM}$ further contributes towards the thermalization energy $E_{\rm th}$ at this time. 
By considering all contributions to $E_{\rm th}$ at $t=16$~ms, including the initial BEC interaction energy, the energy added due to the Bragg diffraction process, an estimated initial temperature of $0.3T_c$, as well energy due to the decay of center-of-momentum oscillations, we expect the system to rethermalize at a temperature of $T=255\pm6$~nK.
Further contributions to the heating of the system could be a result of non-adiabaticity in the initial ramping of the potential, or collisions with background gases.

We find that the growth rate of systems in which a condensed fraction is formed depends strongly on the exact conditions of the system, in particular, the number of atoms within the system, as well as the final temperature at which the system rethermalizes. In order to study the rethermalization rate as a function of trap frequency, we therefore consider systems that rethermalize above the critical temperature $T_c$, such that there is no formation of a condensed fraction. To achieve this, we initially heat the BEC by truncating the evaporation early, such that the Bragg diffraction process provides enough energy to exceed $T_c$ upon rethermalization. We perform this experiment for a range of $\omega_y$, and extract growth rates in the same way as for Fig.~\ref{fig:convFit}(a). The results are shown in Fig.~\ref{fig:convFit}(c) as blue circles, with the growth rate from Fig.~\ref{fig:convFit}(a) also shown as a red cross for reference. We find that the growth rate increases with increasing trapping frequency, to be expected as the shorter oscillation period within the potential results in a greater collision rate between atoms in the $|0\hbar k_\ell\rangle$ and $|2\hbar k_\ell\rangle$ momentum ensembles. The higher growth rate of the systems that thermalize below $T_c$ compared to systems that thermalize above $T_c$ is attributed to the Bose-enhancement that is apparent with the growth of a condensed fraction.

\section{Simulations}\label{sec:simulation}

\begin{figure}[t]
\includegraphics[]{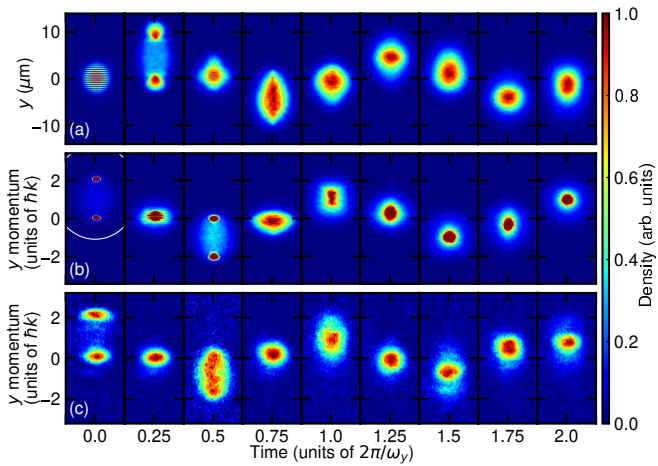}
\caption{\label{f:cloudshots}Snapshots of the system evolution following the Bragg pulse using Eq.~\eqref{e:schro} in (a) real space, integrating over $z$ with horizontal axis $x$, and (b) momentum space, integrating over $k_z$ with horizontal axis $k_x$, with $N=2\times 10^4$ and $(\omega_x,\omega_y,\omega_z)/(2\pi) = (264,220,440)$~Hz and $w=35\:\mu$m. The white circle shown in the first image of (b) indicates the region where noise is added to the truncated Wigner simulation. (c) Experimental time-of-flight images corresponding to the theoretical images. The images selected do not exactly correspond to multiples of $\omega_y/4$ due to the 200~$\mu$s time steps of the experiment. The experiment also involves a 10~ms free expansion during which the system will continue to evolve, hence care must be taken when making comparisons with the momentum images of (b).}
\end{figure}

To numerically demonstrate the thermalization of the average momentum mode after the Bragg pulse, we use the truncated Wigner method which evolves according to the time-dependent Gross-Pitaevskii equation 
\begin{align}
    i\hbar \frac{\partial \psi(\textbf{r},t)}{\partial t} = \left[H_{sp} + g|\psi(\textbf{r},t)|^2 \right]\psi(\textbf{r},t)\label{e:schro}
\end{align}
where
\begin{align}
    H_{sp} \!=\! -\frac{\hbar^2 \nabla^2}{2m} \!+\! \frac{m}2\left[\omega_x^2x^2 + \omega_y^2\frac{w^2}2(1-e^{-2y^2/w^2}) + \omega_z^2 z^2\right]
\end{align}
is the single particle Hamiltonian, $g=4\pi\hbar^2a_s/m$ is the $s$-wave coupling constant, with $a_s=109a_0$ the $s$-wave scattering length and $w$ is the $1/e^2$ radius of the dipole trapping lasers. For $|y|\ll w$, $H_{sp} \approx H^0_{sp} = -\frac{\hbar^2 \nabla^2}{2m} + \frac{m}2(\omega_x^2x^2 + \omega_y^2y^2 + \omega_z^2 z^2)$. We start by finding the ground state $\psi_0(\textbf{r})$ of $[H_{sp}^0 + g|\psi_0|^2]\psi_0(\textbf{r})=\mu\psi_0(\textbf{r})$, with $\mu$ the chemical potential, and calculate the initial state using 
\begin{align}
    \psi(\textbf{r},0) = \frac1{\sqrt2}(1 + e^{2ik_\ell y})\psi_0(\textbf{r}) + e^{ik_\ell y}\sideset{}{'}\sum_n \alpha_n \phi_n(\textbf{r}) \label{e:initnoise} 
\end{align}
where the term $1 + e^{2ik_\ell y}$ creates an equal superposition of wavepackets in momentum 0 and $2\hbar k_\ell  \hat{y}$ states. The $\alpha_n$ are complex Gaussian random variables with $\langle|\alpha_n|\rangle^2 = \frac12$ to account for quantum fluctuations in the truncated Wigner prescription. We add noise in the single-particle basis, i.e. $\phi_n$ are the eigenstates of the single particle Hamiltonian, $H_{sp}^0\phi_n = \epsilon_n\phi_n$. The prime in Eq.~\eqref{e:initnoise} limits the sum to $\epsilon_n < 3\hbar^2k_\ell^2/2m$. We include the factor $e^{ik_\ell y}$ to center the noise about $k_y=k_\ell $ to ensure there is noise covering all modes from below $k_y=0$ to above $k_y=2k_\ell $ without adding excessive noise, which would decrease the time over which the truncated Wigner approximation remains valid. The presented results average over 50 shots and show the symmetrically ordered expectations.

The truncated Wigner simulation is expected to reproduce the quantum evolution of the system over short timescales \cite{Blakie2008a}. Because of the large energy added by the Bragg pulse, the system is highly excited. A large number of spatial modes play a role (the added noise amounts an addition of $\sim 4600$ particles, i.e. an extra 23\% of $N=2\times 10^4$ particles). Due to the quantum degeneracy, the continued center of mass motion, the importance of modes with initially low occupation, and the highly non-equilibrium evolution, few methods other than truncated Wigner are available to model the rethermalization. However, eventually the added noise added itself thermalizes, after which  the numerical results no longer correctly reflect the quantum evolution of the system. Similar calculations were performed for 16\% of a trap period with plane-wave noise in \cite{Norrie2005a} and for 5\% of a trap period using positive-P in \cite{Deuar2007a}, in both cases with the trap removed immediately after the Bragg pulse. 

We integrate over $k_z$ to show the $k_y$ momentum density (vertical) and $k_x$ momentum density (horizontal) in Fig.~\ref{fig:experimentEvolution}(b). Initially there are equal components of the momentum density at $k_y=0$ and $k_y=2k_\ell $ and the development of the $k_y=k_\ell$ mode is seen.

In Fig.~\ref{fig:convFit}(a) we obtain the black curve by integrating the momentum density over a $k_x$, $k_z$ and a $k_y$ window of size $0.45\hbar k_\ell $ about the centre of momentum.

We show the evolution of the system in real space in Fig.~\ref{f:cloudshots}(a) and momentum space in Fig.~\ref{f:cloudshots}(b), along with the corresponding experimental images in Fig.~\ref{f:cloudshots}(c). Initially the condensate is prepared into a superposition of wavepackets with momentum 0 and $2\hbar k_\ell \hat{y}$. The excited wavepacket undergoes oscillatory motion due to the confinement provided by the potential. After half of a trap period, the two wavepackets re-collide in position space and a scattering halo is seen to form in momentum space. As this evolution continues, the system is seen to evolve towards a single peak with a thermal like distribution. This peak continues to oscillate in the trap.

\section{Spontaneous defect formation}\label{sec:sectionfour}
\begin{figure}[t]
	\includegraphics{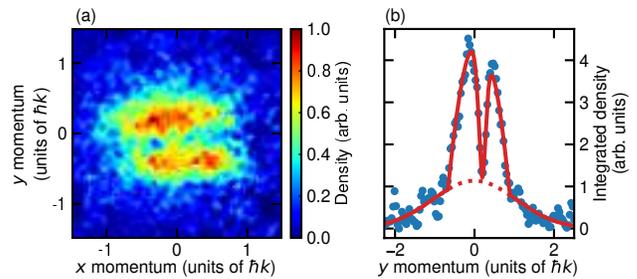}
	\caption{(a) Experimental time-of-flight absorption image featuring a gray soliton of near to 100\% contrast. The image is taken after $t=15$~ms of in-potential evolution with trap frequency $\omega_y/(2\pi) = 280$~Hz. (b) Integrating a strip of the atomic density reveals the nature of the soliton. We perform a fit to the density profile consisting of a Thomas-Fermi profile for the condensed atoms, a Bose-Enhanced Gaussian for the non-condensed atoms~\cite{Szczepkowski2009, Ruddell2015}, as well as a characteristic gray soliton profile. 
		}
	\label{fig:soliton}
\end{figure}

\noindent Following combination of atoms into the $|1\hbar k_\ell\rangle$ ensemble and the subsequent growth of a condensed fraction, the system oscillates within the potential as a mixture of both condensed and non-condensed atoms. In a fraction of our experiments, we observe the presence of defects within the condensed fraction, which we identify as gray solitons~\cite{Lamporesi2013}, as shown in Fig.~\ref{fig:soliton}(a).

As Bose-Einstein condensation takes place, the rate at which information regarding the overall phase of the wavefunction travels is limited by the speed of sound in the condensate. When the front of the condensate transition propagates faster than the speed of sound, phase coherence is not established between multiple regions of condensation, resulting in phase domains differing by $\left|\Delta\phi\right| \leq \pi$. When these condensates approach each other, their boundary forms an unstable gray soliton, a process known as the Kibble-Zurek mechanism~\cite{Zurek1985,Zurek1996,Zurek2009}.

Figure~\ref{fig:soliton}(a) shows a time-of-flight image containing a soliton within the condensed fraction of the atoms, taken after $t = 15$~ms of evolution within the potential, for a harmonic trapping frequency of $\omega_y/(2\pi)=280$~Hz. By integrating a slice of the absorption image in Fig.~\ref{fig:soliton}(a), we can gain insight into the density profile, as shown in Fig.~\ref{fig:soliton}(b). Here we have applied a fit of a Thomas-Fermi profile for the condensed atoms, a Bose-enhanced Gaussian for the non-condensed atoms~\cite{Szczepkowski2009,Ruddell2015}, and a characteristic $\tanh^2\left[(y-y_0)/\sigma_s\right]$ profile for the gray soliton, where $y_0$ is the soliton offset and $\sigma_s$ is related to the soliton width. We perform multiple experiments having $t=15$~ms of evolution, from which we find 20\% of images containing features that we are able to readily identify as solitons.

The soliton contrast in a BEC system is proportional to $1 - v_s/c_s$ \cite{Burger1999}, with $v_s$ the soliton velocity and $c_s$ the speed of sound within the condensed fraction. At non-zero temperature, the soliton interacts with the non-condensed atoms, accelerating and dissipating energy in the form of vortices~\cite{Campo2011}, leading to a reduction in contrast until the soliton vanishes. Although we expect vortices to appear as the solitons decay, we are unable to identify their presence in our time-of-flight images.  

\section{Summary}\label{sec:summary}
We have experimentally and numerically investigated thermalization processes of a Bose gas within a three-dimensional, Gaussian potential. Initially prepared in a highly non-equilibrium state, the system evolves towards equilibrium within the potential, where we observe the initial destruction of the remaining BEC, before the subsequent growth a new condensate fraction in an oscillating reference frame. The growth rate of atomic density in this reference frame is characterized, and we find that it is enhanced compared to experiments where a condensed fraction is not formed. We perform numerical simulations of the system evolution using the truncated Wigner method, and find good agreement with our experimental results for short time scales. Following the growth of a condensed fraction in the oscillating frame, in 20\% of images we observe the presence of solitons within the condensate, which we attribute to the Kibble-Zurek mechanism. Finally, we observe that anharmonicities felt by the atoms as they oscillate away from the trap minima over the timescale of the experiment contribute as much as 20\% of the initially added energy, which would otherwise be locked in the center-of-mass oscillation,  to the energy available for  thermalization. These anharmonicities could possibly be harnessed for designing time-dependent potentials capable of enhancing thermalization rates, allowing for faster and more efficient production of Bose-Einstein condensates.

\begin{acknowledgments}
We would like to thank P.~B.~Blakie, J.~G.~Cosme, and J.~Brand for stimulating discussions. This research was supported by the Marsden Fund, administered by the Royal Society of New Zealand. 
D. B. acknowledges the contribution of NZ eScience Infrastructure (NeSI) high-performance computing facilities.
\end{acknowledgments}

\raggedright

\end{document}